# Cosmic Feedback from Supermassive Black Holes

*Andrew C. Fabian*[1,2,3]

[1] Institute of Astronomy, University of Cambridge, Madingley Road, Cambridge CB3 0HA, United Kingdom

[2] Phone: +011-44-01223-37509

[3] E-mail: acf@ast.cam.ac.uk

E. Churazov[4], M. Donahue[5], W. R. Forman[6], M. R. Garcia[6], S. Heinz[7], B. R. McNamara[8], K. Nandra[9], P. Nulsen[6], P. Ogle[10], E. S. Perlman[11], D. Proga[12], M. J. Rees[1], C. L. Sarazin[13], R. A. Sunyaev[4], G. B. Taylor[14], S. D. M. White[4], A. Vikhlinin[6], D. M. Worrall[15]

4) Max-Planck-Institut für Astrophysik, Karl Schwarzschild Str. 1, Garching, Germany

5) Department of Physics and Astronomy, Michigan State University, East Lansing, MI 48824

6) Harvard-Smithsonian Center for Astrophysics, 60 Garden St., Cambridge MA, 02138

7) Department of Astronomy, University of Wisconsin-Madison, 475 N. Charter St., Madison, WI 53706-1582

8) University of Waterloo, Physics & Astronomy, 200 University Avenue W., Waterloo, Ontario, N2L 3G1; Perimeter Institute for Theoretical Astrophysics, Waterloo, Ont.

9) Astrophysics group, Imperial College, Prince Consort Road London SW7 2AZ, United Kingdom

10) Spitzer Science Center, California Institute of Technology, Pasadena, CA 91125

11) Department of Physics and Space Science, Florida Institute of Technology, 150 W. University Blvd, Melbourne, Florida 32901

12) Department of Physics and Astronomy, University of Nevada-Las Vegas, 4505 Maryland Parkway, Las Vegas, NV 89154-4002

13) Department of Astronomy, University of Virginia, P.O. Box 400325, Charlottesville, VA 22904

14) Department of Physics and Astronomy, University of New Mexico, MSC 07 4220, Albuquerque, NM 87131

15) Department of Physics, University of Bristol, Tyndall Ave., Bristol BS8 1TL, United Kingdom

# Cosmic Feedback from Supermassive Black Holes

An extraordinary recent development in astrophysics was the discovery of the fossil relationship between central black hole mass and the stellar mass of galactic bulges. The physical process underpinning this relationship has become known as feedback. The *Chandra* X-ray Observatory was instrumental in realizing the physical basis for feedback, by demonstrating a tight coupling between the energy released by supermassive black holes and the gaseous structures surrounding them. A great leap forward in X-ray collecting area and spectral resolution is now required to address the following question:

*How did feedback from black holes influence the growth of structure?*

## Feedback and galaxy formation

Every massive galaxy appears to have a massive black hole at its center whose mass is about 0.2% of the mass of the galaxy's bulge (Tremaine et al. 2002). It is now widely considered that the black hole may have brought about this correlation by regulating the amount of gas available for star formation in the galaxy. Massive black holes thereby have a profound influence on the evolution of galaxies, and possibly on their formation. Several puzzling aspects of galaxy formation, including the early quenching of star formation in galactic bulges and the galaxy mass function at both high and low ends, have been attributed to black hole "feedback" (Croton et al. 2006).

In size, a black hole is to a galaxy roughly as a person is to the Earth. Something very small is determining the growth of something very large. This is possible because the gravitational potential energy acquired by an object approaching a black hole is a million times larger than the energy of an object orbiting in the potential of a typical galaxy. As a black hole grows to 0.2% of the bulge mass through accreting matter, it releases nearly 100 times the gravitational binding energy of its host galaxy.

There is no question that a growing black hole *could* drastically affect its host galaxy. Whether and how it *does* so, however, is an open question that depends on how much of the energy released actually interacts with the matter in the galaxy. If the energy is in electromagnetic radiation and the matter largely stars, then very little interaction is expected. If the matter consists of gas, perhaps with embedded dust, the radiative output of the black hole can both heat the gas, and drive it via radiation pressure. Alternatively, if significant AGN power emerges in winds or jets, mechanical heating and pressure provide the link. Either form of interaction can be sufficiently strong that gas can be driven out of the galaxy entirely (Silk & Rees 1998).

The radiative form of feedback is most effective when the black hole is accreting close to its Eddington limit. The mechanical form associated with jets, on the other hand, operates at rates below the Eddington limit. X-ray observations are essential for studying both forms of feedback. The mechanical forms of feedback rely on dynamical (ram) pressure to accelerate gas to high speeds (Fig. 1). If this gas is initially of moderate temperature, the interaction will shock it to high temperatures where it can only be detected in X-rays. Since the efficiency of ram pressure acceleration is roughly proportional to the volume filling factor of the accelerated gas, most of the energy is probably absorbed by the hot component of the interstellar medium in any case. Gas accelerated by radiation pressure or radiative heating is likely to be cold and dusty. The interaction is therefore much more difficult to observe directly. X-ray and far infrared emission can emerge from the inner regions where the interaction occurs, revealing the Active Galactic Nucleus (AGN) itself.

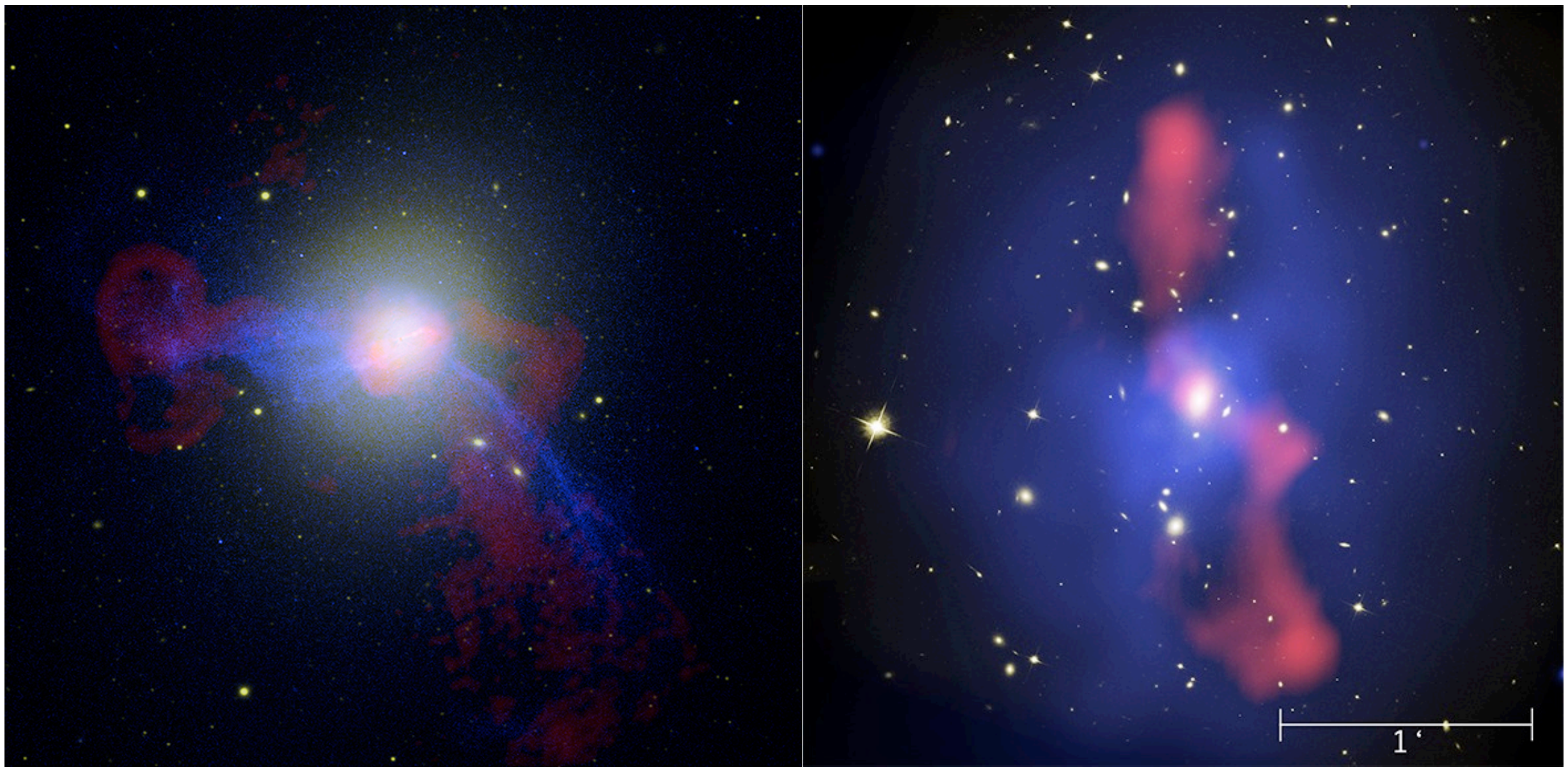

**Fig. 1, Left:** X-ray emission (blue), radio emission (red) superposed on an optical image of M87. The X-ray structure (shocks, bubbles) was induced by a ~$10^{58}$ erg outburst that began 10 Myr ago (Forman et al. 2005). The persistence of the delicate, straight-edge X-ray feature indicates a lack of strong turbulence. We expect IXO to reveal ordered velocity structure. The image is 50 kpc on a side. **Right:** X-ray emission (blue), 320 MHz radio emission (red) superposed on an HST image of the z=0.21 cluster MS0735.6+7421 (McNamara et al. 2005). The image is 700 kpc on a side. Giant cavities, each 200 kpc (1 arcmin) in diameter, were excavated by the AGN. The mechanical energy is reliably measured in X-rays by multiplying the gas pressure by the volume of the cavities, and by the properties of the surrounding shock fronts. With a mechanical energy of $10^{62}$ erg, MS0735 is the most energetic AGN known. This figure shows that AGN can affect structures on galaxy scales of tens of kpc and on cluster-wide scales, spanning hundreds of kpc in MS0735.

Radiative acceleration appears to be particularly dramatic in the outflows from some luminous quasars. UV observations indicate that outflows reaching 0.1-0.4c may be present in most quasars. X-ray observations are required to determine the total column density and hence the kinetic energy flux. Current work on a small number of objects

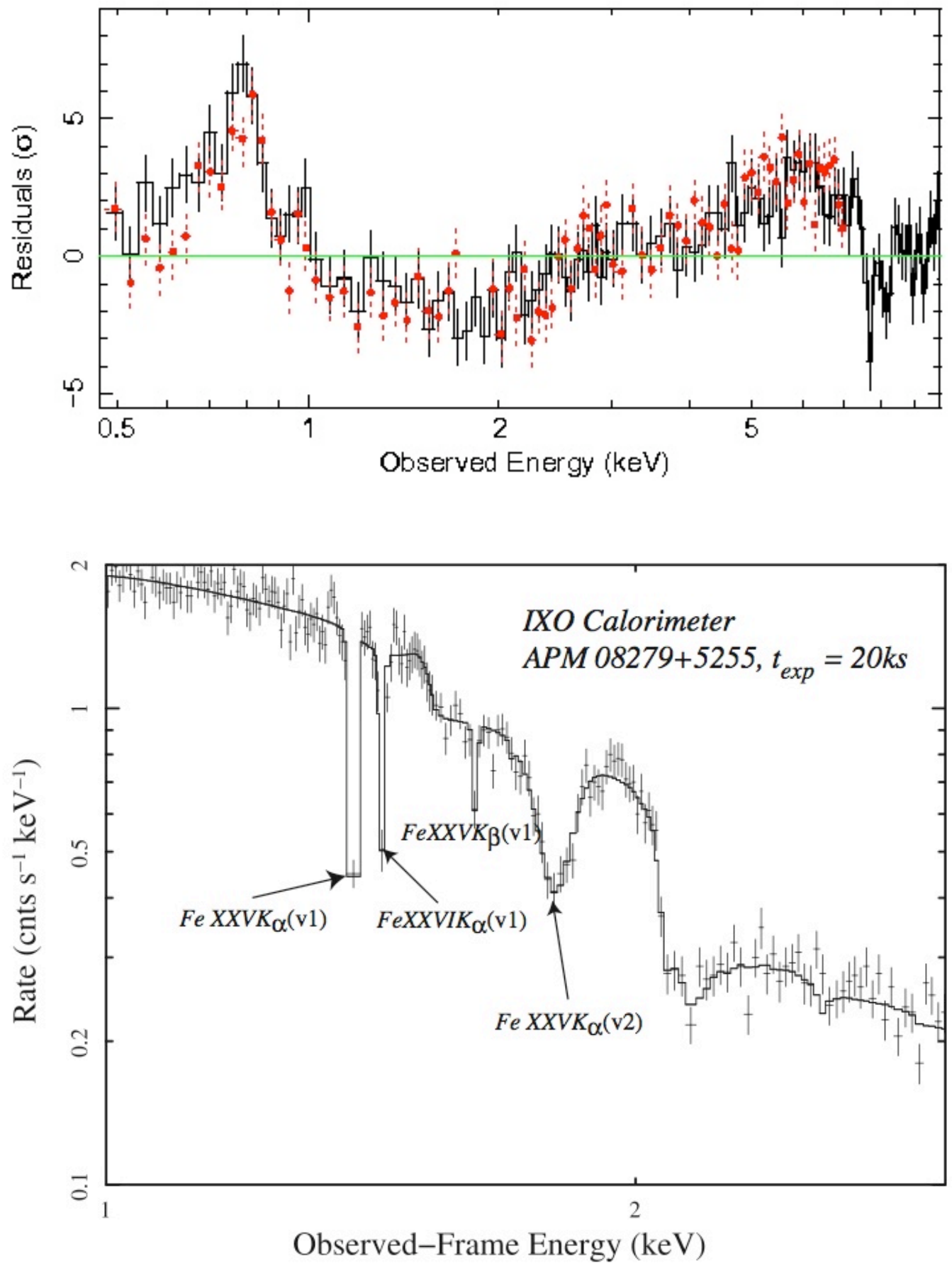

**Fig. 2, Top:** *Suzaku* X-ray spectrum of luminous quasar PDS456 at z=0.184. The strong iron absorption lines at 9 keV in the rest frame indicate an outflow at 0.3c and a mechanical power of $10^{47}$ erg $s^{-1}$ (Reeves et al. 2009).

**Bottom:** 20 ksec simulation of an IXO spectrum of the absorption structure of the mildly relativistic outflow seen by *Chandra* in the BAL quasar APM08279 at z=3.91.

implies that this can be comparable to the radiative luminosity (Fig. 2). To diagnose the energetics of quasars we need large samples of quasars, comparing them with the less energetic (but still substantial) outflows from lower luminosity AGN, which can reach speeds of several thousand km/s.

IXO's huge spectroscopic throughput will extend this to redshifts *z*=1-3, where the majority of galaxy growth is occurring. IXO will be sensitive to all ionization states from Fe I – Fe XXVI, allowing us to study how feedback affects all phases of interstellar and intergalactic gas, from million-degree collisionally ionized plasmas to ten-thousand degree photoionized clouds. These measurements will probe over 10 decades in radial scale, from the inner accretion flow where the outflows are generated to the halos of galaxies and clusters, where the outflows deposit their energy.

Current models of galaxy evolution predict that merger-induced star formation and AGN activity proceeds under heavy obscuration, building the galaxy's bulge and black hole before the AGN blows out all of the gas and terminates star formation. Following this, an unobscured QSO is briefly revealed before the galaxy becomes much less active as a massive red elliptical. The complete census of AGN enabled by IXO will pinpoint galaxies whose black holes are undergoing all these phases of evolution, and crucially the heavily obscured "Compton thick" phase predicted during the quenching of star formation. Observations of galaxies in the *X-ray band* are a powerful way to select accreting black holes in an unbiased fashion, and to probe the inner workings of AGN near the black hole's gravitational radius.

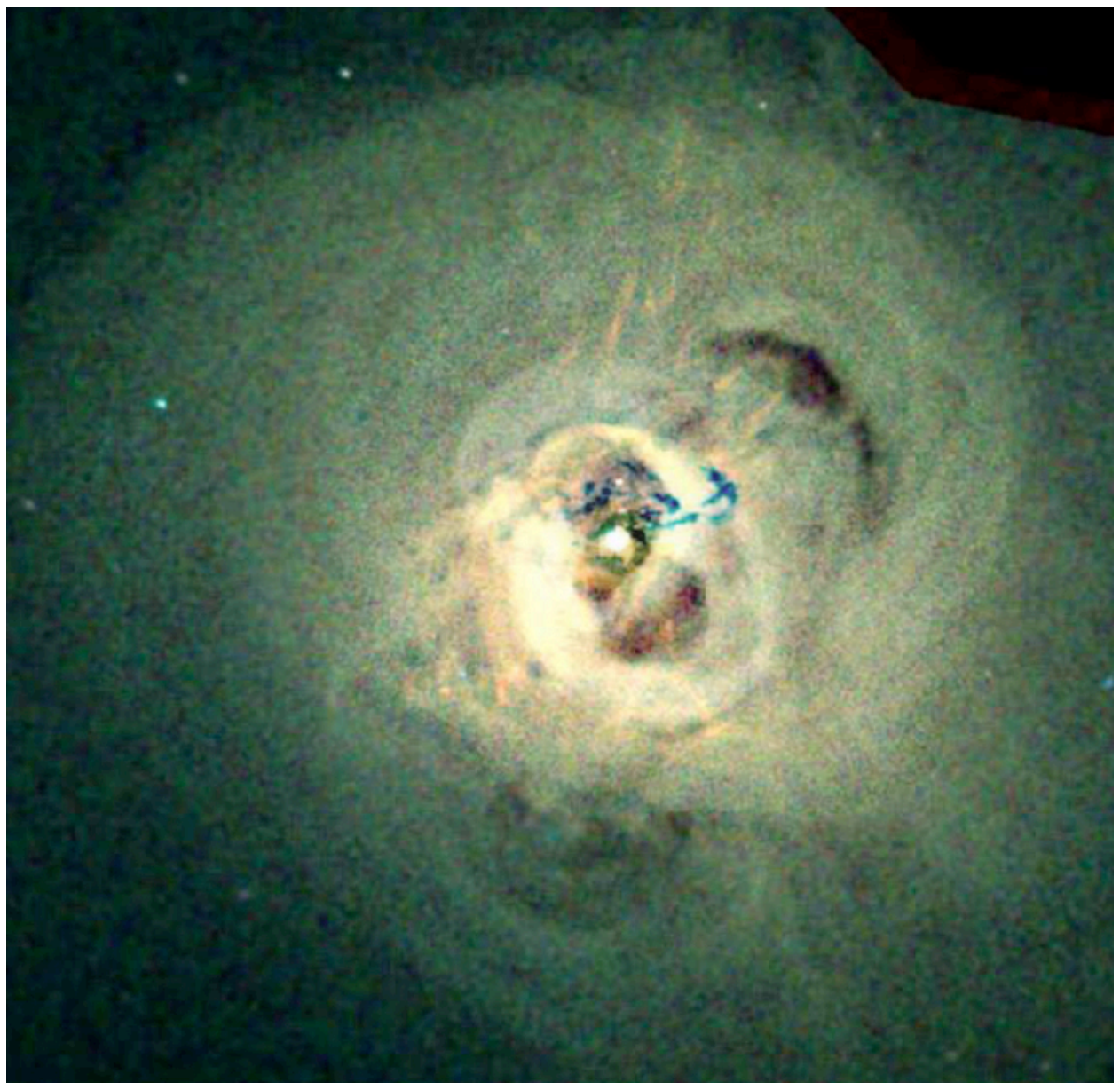

**Fig. 3:** How did feedback from black holes influence galaxy growth? *Chandra* X-ray observations of Perseus (left; Fabian et al. 2003) and other nearby clusters have revealed the indelible imprint of the AGN on the hot gas in the core. Radiative and mechanical heating and pressure from black holes have a profound influence not only on the hot baryons, but on the evolution of all galaxies whether or not they are in clusters.

## Thermal regulation of gas in bulges and cluster cores

Mechanical feedback dominates in galaxies, groups, and clusters at late times, as shown by X-ray observations of gas in the bulges of massive galaxies and the cores of galaxy clusters (eg., Fig. 1). The energy transfer process is surprisingly subtle. The radiative cooling time of the hot gas in these regions is often much shorter than the age of the system, so that without any additional heating, the gas would cool and flow into the center. For giant ellipticals the resulting mass cooling rates would be of order 1 solar mass per year. At the centers of clusters and groups, cooling rates range between a few to thousands of $M_\odot$ per year. Spectroscopic evidence from *Chandra* and *XMM* shows that some cooling occurs, but not to the extent predicted by simple cooling (Peterson & Fabian 2006). Limits on cool gas and star formation rates confirm this. Mechanical power from the central AGN acting through jets must be compensating for the energy lost by cooling across scales of tens to hundreds of kpc (McNamara & Nulsen 2007).

The gross energetics of AGN feedback in galaxies and clusters are reasonably well established (Fig. 4). Remarkably, relatively weak radio sources at the centers of clusters often have mechanical power comparable to the output of a quasar, which is sufficient to prevent hot atmospheres from cooling (McNamara & Nulsen 2007). The coupling between the mechanical power and the surrounding medium are, however, poorly understood. Moreover, it is extremely hard to understand how a fine balance can be established and maintained.

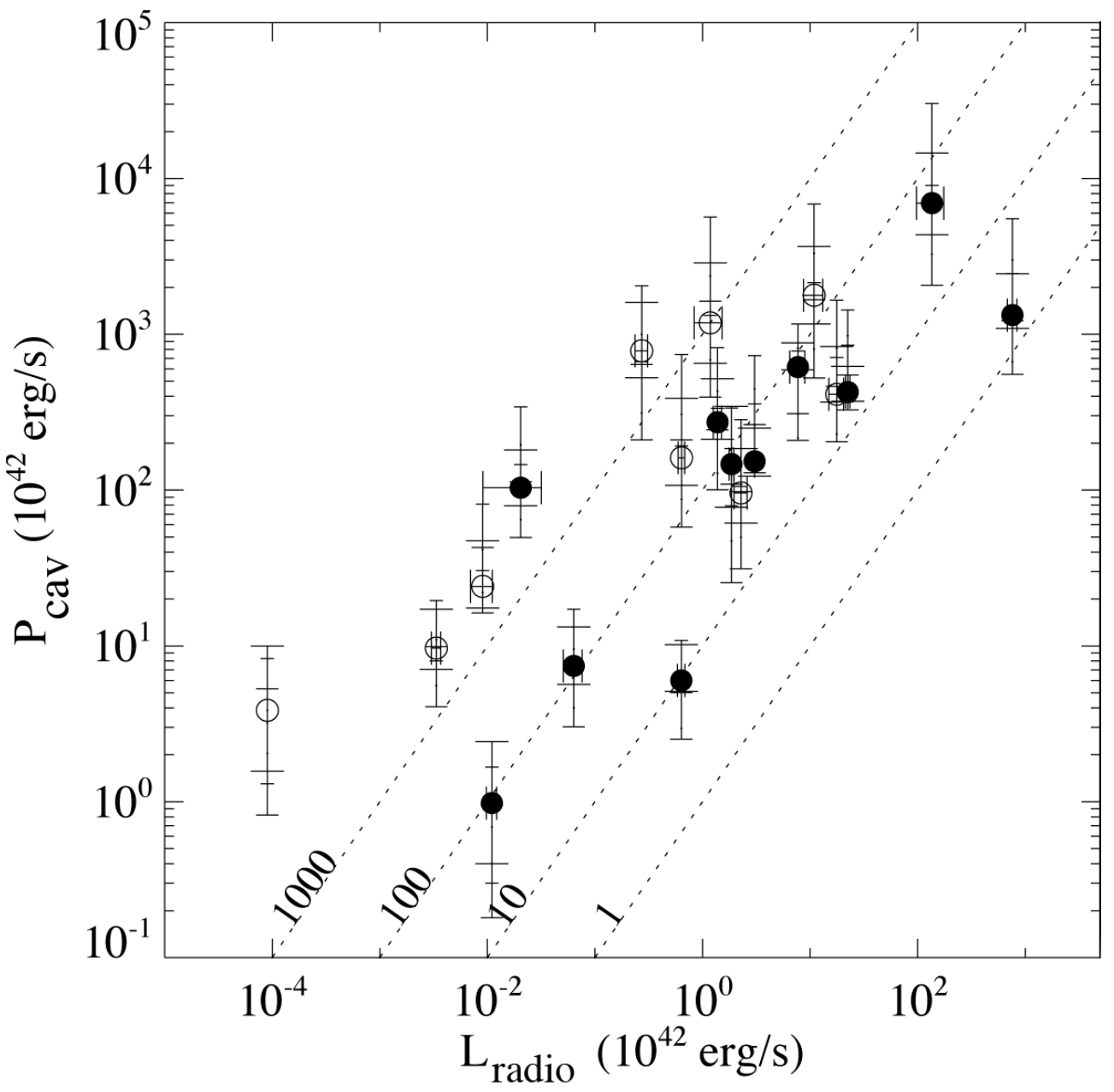

**Fig. 4:** Jet power inferred from X-ray cavities in clusters of galaxies (McNamara & Nulsen 2007) plotted against radio luminosity (10Mhz – 10GHz). The diagonal lines represent ratios of constant jet power to radio synchrotron power. Jet power correlates with synchrotron power but with a *very* large scatter in their ratio. Radio sources embedded in hot atmospheres are dominated by mechanical power.

The heat source – the black hole – is roughly the size of the Solar System, yet the heating rate must be tuned to conditions operating over scales 10 decades larger. The short radiative cooling time of the gas means that the feedback must be more or less continuous. How the jet power, which is highly collimated to begin with, is isotropically spread to the surrounding gas is not clear. The obvious signs of heating include bubbles blown in the intracluster gas by the jets (Figs. 1, 3) and nearly quasi-spherical ripples in the X-ray emission that are interpreted as sound waves and weak shocks. Future low-frequency radio observations of the bubbles and cavities are of great importance in determining the scale of the energy input. The disturbances found in the hot gas carry enough energy flux to offset cooling, but the microphysics of how such energy is dissipated in the gas is not understood.

The persistence of steep abundance gradients in the cluster gas, imprinted by supernovae in the central galaxy means that the feedback is gentle, in the sense that it does not rely on violent shock heating or supersonic turbulence. Long filaments of optical line-emitting gas in some objects suggest low levels of turbulence. Yet the continuous streams of radio bubbles made by the jets, the movement of member galaxies and occasional infall of subclusters must make for a complex velocity field.

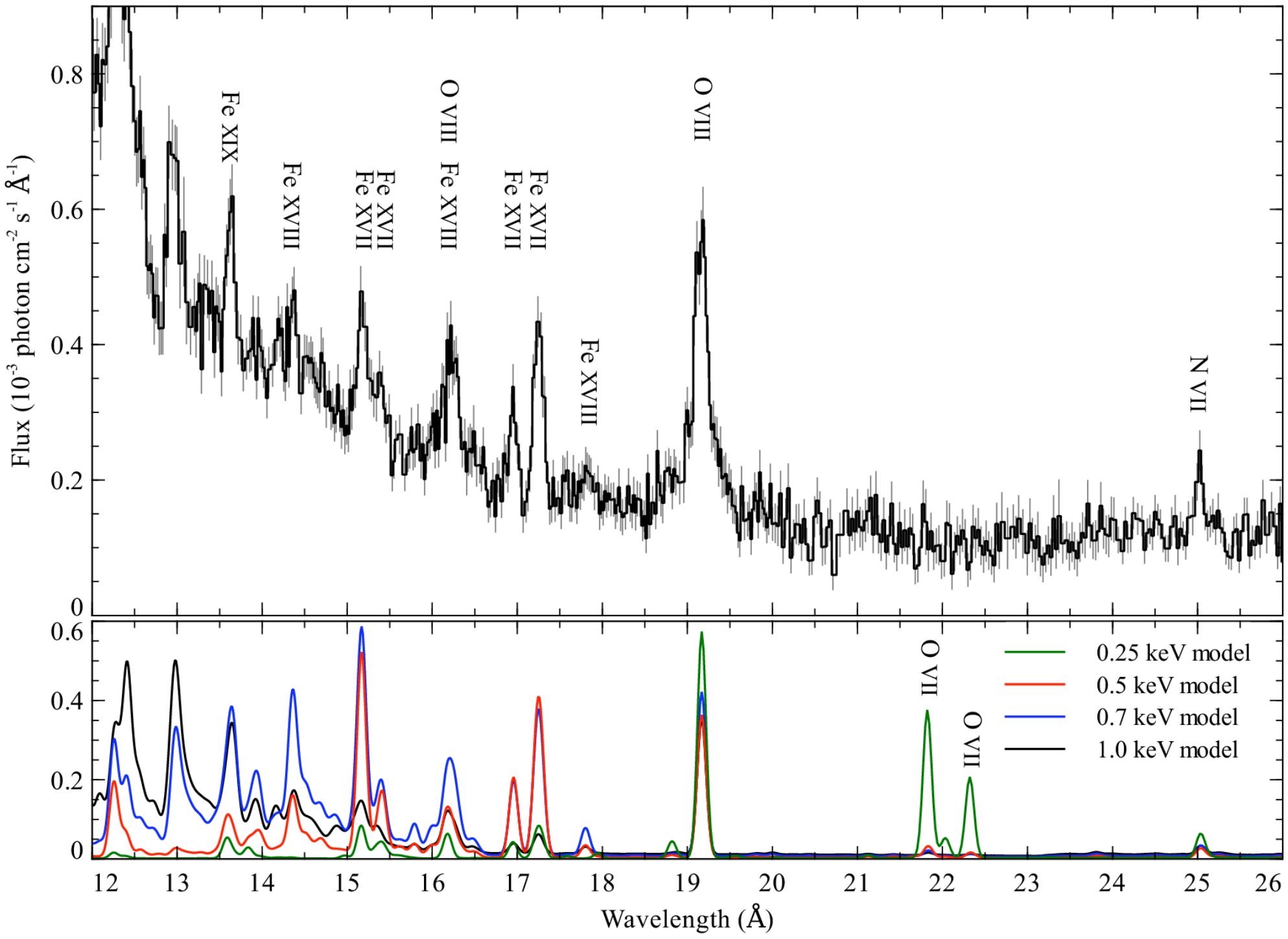

**Fig. 5:** *XMM-Newton* reflection grating spectrometer spectrum of the core of the Centaurus cluster (Sanders et al. 2008). This part of the spectrum shows a rich set of emission lines from oxygen, neon, and iron. The ionization states of iron show the temperature structure of the gas ranging from 0.4 keV to 4 keV.

With high resolution imaging and moderate resolution spectroscopy (Fig. 5), the *Chandra* and *XMM-Newton* observatories have established AGN feedback as a fundamental astrophysical process in nature. However, the dynamics of these powerful outflows are not understood.

This demands a leap in spectral resolution by one to two orders of magnitude above that of *Chandra* and *XMM-Newton*. The spectral resolution and sensitivity of the next generation X-ray observatory, such as IXO, is needed to understand how the bulk kinetic energy is converted to heat. Its capabilities are essential in order to measure and map the gas velocity to an accuracy of ten km/s, revealing how the mechanical energy is spread and dissipated (Fig. 6). From accurate measurements of line profiles and from the variations of the line centroid over the image it is possible to deduce the characteristic spatial scales and the velocity amplitude of large (> kpc) turbulent eddies, while the total width of the line provides a measure of the total kinetic energy stored in the stochastic gas motions at all spatial scales. Such data will provide crucial insight into the ICM heating mechanisms. Observations of the kinematics of the hot gas phase, which contains the bulk of the gaseous mass, and absorbs the bulk of the mechanical energy in massive elliptical galaxies, are only possible at X-ray wavelengths.

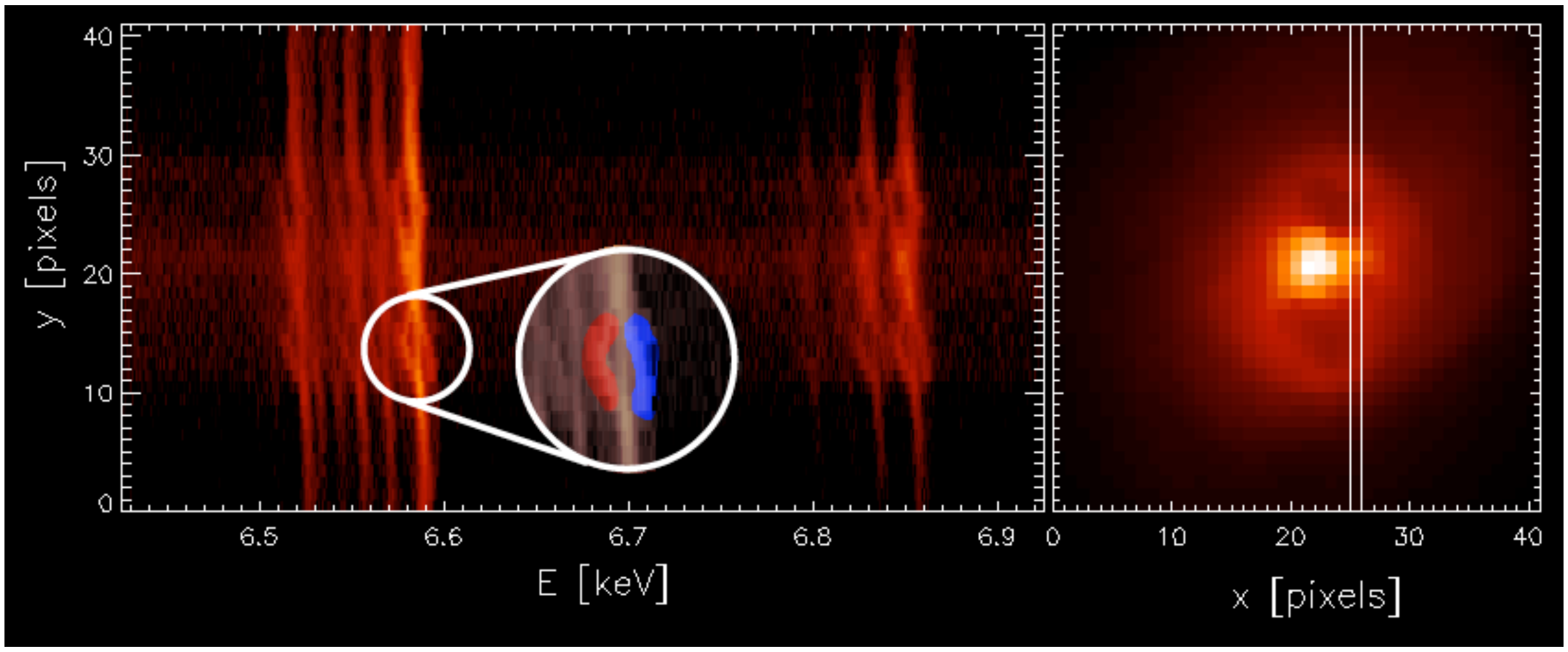

**Fig. 6.** Simulated high-resolution X-ray spectra from the shells and X-ray cavities in the Perseus cluster (see Fig. 3), demonstrating the power of imaging spectroscopy for AGN feedback studies. The right panel shows the X-ray image and the chosen cut (spectral slit, slicing through both cavities), while the left shows the spectrum of the K-alpha lines from Fe XXV and Fe XXVI (both lines are multiplets), as would be observed by the IXO micro calorimeter spectrograph in an exposure of 250,000 seconds. At the location of the cavities (y=10-15 and y=25-30), each of the lines splits into three components: The front wall of the cavity (blue-shifted), the rear wall of the cavity (red-shifted), and a rest-frame component from fore- and background cluster emission. The expansion velocity and thus the age of the cavity can easily be determined from the red- and blue shifts of the lines, allowing a precise determination of the jet power. The cluster/radio galaxy model used for the spectral simulations is the result of direct hydrodynamic simulations of jets in galaxy clusters with parameters appropriate for Perseus (jet power $10^{45}$ erg $s^{-1}$; Heinz et al. 2009, in preparation). See www.astro.wisc.edu/~heinzs/perseus for a movie.